\documentclass[german,english,aps,preprint]{revtex4}
\usepackage[latin9]{inputenc}
\usepackage{float}
\usepackage{bm}
\usepackage{amsmath}
\usepackage{amssymb}
\usepackage{graphicx}
\usepackage{esint}

\makeatletter

\@ifundefined{textcolor}{}
{%
 \definecolor{BLACK}{gray}{0}
 \definecolor{WHITE}{gray}{1}
 \definecolor{RED}{rgb}{1,0,0}
 \definecolor{GREEN}{rgb}{0,1,0}
 \definecolor{BLUE}{rgb}{0,0,1}
 \definecolor{CYAN}{cmyk}{1,0,0,0}
 \definecolor{MAGENTA}{cmyk}{0,1,0,0}
 \definecolor{YELLOW}{cmyk}{0,0,1,0}
}

\makeatother

\usepackage{babel}
\begin{document}

\preprint{This line only printed with preprint option}

\title{Dealing with the exponential wall in electronic structure calculations}

\author{Peter Fulde}
\email{fulde@pks.mpg.de}
\affiliation{Max-Planck-Institut f\"ur Physik komplexer Systeme, N\"othnitzer Stra\ss e 38, 01187 Dresden, Germany}

\author{Hermann Stoll}
\email{stoll@theochem.uni-stuttgart.de}
\affiliation{Institut für Theoretische Chemie, Universit\"at Stuttgart, Pfaffenwaldring 55, 70550 Stuttgart, Germany}

\date{\today}
\begin{abstract}
An alternative to Density Functional Theory are wavefunction based electronic structure calculations for solids. In order to perform them the Exponential Wall (EW) problem has to be resolved. It is caused by an exponential increase of the number of configurations with increasing electron number N. There are different routes one may follow. One is to characterize a many-electron wavefunction by a vector in Liouville space with a cumulant metric rather than in Hilbert space. This removes the EW problem. Another is to model the solid by an {\it impurity} or {\it fragment} embedded in a {\it bath} which is treated at a much lower level than the former. This is the case in Density Matrix Embedding Theory (DMET) or Density Embedding Theory (DET). The latter are closely related to a Schmidt decomposition of a system and to the determination of the associated entanglement. We show here the connection between the two approaches. It turns out that the DMET (or DET) has an identical active space as a previously used Local Ansatz, based on a projection and partitioning approach. Yet, the EW problem is resolved differently in the two cases. By studying a $H_{10}$ ring these differences are analyzed with the help of the method of increments.
\end{abstract}
\maketitle

\section{Introduction}
\label{Introduction}

Precise electronic structure calculations for periodic systems are an important field of research in the theory of condensed matter.  In approaches like Density Functional Theory (DFT) it is not necessary to know, e.g., the many-body ground state wavefunction in order to obtain quantitative answers for ground-state properties such as the lattice constant, binding energy, magnetization etc. Calculations of this type have revolutionized the field. Yet, alternatively one might want to calculate instead the many-body wavefunction and from it various physical properties by applying quantum chemical techniques. In distinction to DFT they allow for controlled approximations and are of interest when results from DFT are  unsatisfactory. This is often the case when electronic correlations are strong. 

Calculating the wavefunction of an interacting electron system faces, however, the so-called Exponential Wall (EW) problem, i.e., that the number of different configurations is exponentially increasing with the electron number. As W. Kohn has pointed out the concept of characterizing a wavefunction by a vector in Hilbert space loses its meaning for electron numbers $N$ larger than approximately $10^3$ \cite{Kohn99}. The high dimension of the configuration space implies that in this case the overlap between any approximate form of the ground-state wavefunction with the exact one is zero for all practical purposes. In order to perform electronic structure calculations based on wavefunctions for large molecules or solids it is therefore mandatory to circumvent the EW problem. The simplest way out is to limit oneself to a self-consistent field (SCF) or Hartree-Fock approximation. The ground-state wavefunction has here the form of a Slater determinant and hence consists of a single configuration, independent of $N$. However, calculations at this level yield results of low quality for various physical quantities. Density functional theory is likewise unaffected by the EW problem \cite{Kohn99,Hohenberg64, Kohn65} since this approach avoids making any statements about the many-electron wavefunction. Instead it is calculating ground-state properties directly from the solutions of the Kohn-Sham equations and it is therefore also of a mean-field type. However, in distinction to the SCF theory it contains correlation effects. They enter through the form of the chosen self-consistent potential in the Kohn-Sham equations. There exists a number of approaches which are in-between a mean-field and a many-body wavefunction. For example, by subdividing a solid into small units one determines the many-electron ground state for each of them, thereby applying periodic boundary conditions. This is done by using, e.g., Coupled Cluster theory \cite{Cizek69,Kuemmel78,Bishop91,Sun99}, perturbation theory (MP2), quantum Monte Carlo (QMC) or full CI-QMC \cite{Manby11,Booth13,Sharma15,Kersten16}. In this case the number of configurations grows exponentially only with the electron number contained in a small unit. The total ground-state wavefunction of the solid is then given by the (antisymmetric) product of the many-electron wavefunctions of the small units. The last step has mean-field character and neglects, e.g., interactions and correlations between electrons in the small subunits. Another way of circumventing the EW problem is by treating the $N$ electron system in form of an impurity model \cite{Georges96,Knizia13,Bulik14}. Thereby a site or a cluster of sites labeled $I$ is considered as an {\it impurity} or a {\it fragment} embedded in an {\it environment} or {\it bath}. The electrons on cluster $I$ are treated on a post-SCF level in the presence of the embedding surroundings, e.g., by exact diagonalization. However, the bath is treated on a much lower level, i.e., a self-consistent field (SCF) level. Thus instead of the total electron number $N$, only the number of electrons on the cluster $I$ is relevant for the dimension of the Hilbert-, or configuration space. The density matrix embedding theory (DMET) \cite{Knizia13} or the density embedding theory (DET) \cite{Bulik14} serve as examples here. Yet, even when one has solved the impurity problem adequately, one still is not able to write down the ground-state wavefunction of the extended system. 

A proper way of avoiding the EW problem is by characterizing the wavefunction of a $N$ electron system not by a vector in Hilbert space but instead by a vector in operator- or Liouville space. In that space a cumulant metric has to be applied \cite{Fulde95_12,Kladko98}. This has been reemphasized in a recent publication \cite{Fulde16} and explained by a particularly simple example. The cumulant metric frees the wavefunction from an exponential large number of configurations, which are redundant and have no effect on physical quantities. It serves a similar purpose as using connected diagrams only in Green's function calculations. Disconnected diagrams drop out at the end of such calculations and therefore are disposed off from the very beginning. Similarly, cumulants ensure that irrelevant, statistically independent terms in the wavefunction do not appear. Coupled Cluster theory \cite{Cizek69,Kuemmel78,Bishop91,Sun99} can be formulated as a special case of the above considerations \cite{Schork92}.

The purpose of this communication is to show the relation between the DMET/DET and the Liouville space approach which differ in the way the EW problem is handled. We also show that the Schmidt decomposition, i.e., the starting point of the DMET/DET, has a one to one correspondence to the projection and partitioning method originally stressed by L\"owdin \cite{Loewdin63,Loewdin86} and elaborated on in the Local Ansatz approach \cite{Fulde95_12,Stollhoff80,Fulde92}.

The paper is divided as follows. In the next Section, we define the Hamiltonian and recall how the EW problem is avoided by defining wavefunctions in operator space with cumulant metric. In Section \ref{ImpEmbMeth} it is shown how impurity embedding theories deal with the EW problem. Section \ref{ImpBeddingCumul}  establishes a connection between the cumulant formulation and the impurity theories. Also numerical comparisons between the two approaches are made.The conclusions and a summary are contained in Section \ref{Concl_Summary}.

\section{Formulation of wavefunctions in Liouville space}
\label{FormWaveLiouSpace}

We start with electronic creation and annihilation operators $a^+_{i \sigma}$ and $a_{i \sigma}$ based on mutually orthogonalized atomic spin-orbitals $f_\nu ({\bf r} - {\bf R}_i) \sigma$ on a lattice. The compact index $i$ includes a site index $I$ and an orbital index $\nu(I)$ with $\nu = 1, \dots, M$. The Hamiltonian is of the standard form
\begin{equation}
H = \sum_{ij \sigma} t_{ij} a^+_{i \sigma} a_{j \sigma} + \frac{1}{2} \sum_{ijkl \atop \sigma \sigma'} V_{ijkl} a^+_{i \sigma} a^+_{k \sigma'} a_{l \sigma'} a_{j \sigma}~~~.
\label{eq:01}
\end{equation}
We decompose $H = H_0 + H_1$ and assume that the ground-state wavefunction $| \Phi_0 \rangle$ of $H_0$ is known. Often $H_0$ will be the self-consistent field part $H_{\rm SCF}$ of $H$ with $| \Phi_0 \rangle$ being the SCF- or HF-ground state. The remaining part $H_1 = H - H_0$ contains the residual interactions. Yet, $H_0$ may also be the Kohn-Sham Hamiltonian $H_{KS}$ with $H_1 = H - H_{KS}$, and a ground state $| \Phi_{KS} \rangle$ in form of a Slater determinant built from Kohn-Sham orbitals. In the following we will limit ourselves to the first, i.e., SCF case. Other choices of $H_0$ can be treated in complete analogy.

The exact ground state $| \psi_0 \rangle$ can be obtained from $| \Phi_0 \rangle$ by means of the wave- or M{\o}ller operator $\tilde{\Omega}$ so that $| \psi_0 \rangle = \tilde{\Omega} | \Phi_0 \rangle$. Because of the EW problem this transformation makes sense only for small electron numbers $N < 10^3$. For $N \gtrsim 10^3$ the number of configurations is so large that the overlap of any approximate wavefunction with the exact one is zero for all purposes. Therefore one has to remove the statistically independent contributions which are responsible for the exponential growth of the number of configurations. Thus the use of cumulants is required. This is seen as follows. Electron correlations affect the surroundings of an electron only up to a certain radius $R_c$. They generate a correlation hole, i.e., they reduce the probability of finding other electrons nearby. Beyond $R_c$ correlation effects may be neglected due to their smallness. The actual size of $R_c$ depends, of course, on the required computational  accuracy. Correlations of electrons farther apart than $2R_c$ are therefore independent of each other for all purposes. Cumulants of matrix elements eliminate statistically independent or factorizable contributions to them. In practice this simply implies taking only connected contractions of creation and destruction operators when the matrix elements are evaluated. We can eliminate the EW problem by characterizing the correlated ground state by the set of operators $A_\mu$ which, upon application on $| \Phi_0 \rangle$, generate the correlations in the electron system \cite{Fulde16}. Thus the operator- or Liouville space spanned  by the $A_\mu$, is used instead of the Hilbert space in order to describe the ground state. The cumulant metric used in Liouville space is introduced in the form 
\begin{equation}
\left( A | B \right) = \langle \Phi_0 \left| A^+B \right| \Phi_0 \rangle^c~~~.
\label{eq:02}
\end{equation}
The upper script refers to taking the cumulant. Cumulants were first used in the free energy expansion of a classical imperfect gas \cite{Mayer40}. Their usefulness in quantum statistical mechanics was pointed out by Kubo \cite{Kubo66}. In \cite{Kladko98} the most important rules for cumulants can be found. 

Of special interest is the transformation behavior when an expression of the form $\langle \Phi_0 | A^+ | \Phi_0 \rangle^c$ is changed into the form $\langle \Phi_0 | A^+ | \psi_0 \rangle^c$. By a sequence of infinitesimal transformations which take us from $| \Phi_0 \rangle$ to $| \psi_0 \rangle$ we can write
\begin{eqnarray}
\langle \Phi_0 \left| A^+ \right| \psi_0 \rangle^c & = & \langle \Phi_0 \left| A^+ \Omega \right| \Phi_0 \rangle^c\nonumber \\
& = & \left( A | \Omega \right)
\label{eq:03}
\end{eqnarray}
where $\Omega$ is the sum of the infinitesimal transformations \cite{Fulde95_12}. We identify $| \psi_0 \rangle$ with the exact ground state of $H$ in which case $| \Omega ) = |1 + S )$ is a cumulant wave operator in analogy to the wave- or M{\o}ller operator \cite{Kladko98}. The cumulant scattering operator $| S )$ is the product of the above mentioned infinitesimal transformations. Note that $| \Omega )$ (or $| S )$) is not unique since several paths in operator space can connect the correlated and uncorrelated ground state $| \psi_0 \rangle$ and $| \Phi_0 \rangle$. When $| \Omega_1 )$ and $| \Omega_2 )$ are two such paths $(A | \Omega_1 - \Omega_2) = 0$ in all cases \cite{Kladko98}. The correlation energy $E_{\rm corr}$ can be written as
\begin{equation}
E_{\rm corr} = (H | S)~~~.
\label{eq:04}
\end{equation}
In passing we mention that the Coupled Cluster theory \cite{Cizek69,Kuemmel78,Bishop91,Sun99} can be written as a special form of $\Omega$, i.e., $| \Omega) = | e^{\tilde{S}})$. Here $\tilde{S}$ is a sum of so-called prime operators, i.e., operators which are not broken up when the cumulant is formed \cite{Schork92}.

By defining the wavefunction by a vector in Liouville space the EW problem has disappeared as explained in more detail in \cite{Fulde16}. In the incremental method \cite{Stoll92,Stoll92a} the cumulant scattering operator $|S)$ is decomposed into one-, two- and multisite contributions
\begin{equation}
| S) = \left( \sum_I S_I + \sum_{\langle IJ \rangle} \delta S_{IJ} + \sum_{\langle IJK \rangle} \delta S_{IJK} + \dots \right)~~~,
\label{eq:05}
\end{equation}
where $IJK$ are site indices and $\delta S_{IJ} = S_{IJ} - S_I - S_J$. Each increment to $|S)$ involves a small number of electrons only. The various increments in (\ref{eq:05}) are computed and the incremental contributions to the correlation energy follow from (\ref{eq:04}) by restricting the annihilation operators in $S_I$ to electrons in spin orbitals centered on site $I$. Similarly these in $S_{IJ}$ are restricted to electrons in spin orbitals centered at site $I$ or $J$ and so on. The different contributions to $| S )$ are computed by standard quantum chemical methods (see e.g. \cite{Paulus06}). Usually the expansion is rapidly convergent. Often, 3-body increments are already small and 4-body increments can be safely neglected. At this stage a comment is in order concerning the type of spin orbitals which are used when the various scattering operators are determined. Choices are localized and local orbitals.

{\it Localized} or Wannier orbitals with creation operators $c^+_{m \sigma} (I)$ are obtained through rotations in the space of occupied-, and (separately) unoccupied Bloch orbitals in $|\Phi_{\rm SCF} \rangle$. We assume band fillings of more than one half, otherwise hole orbitals are used. In terms of the $c^+_{m \sigma} (I)$ the SCF ground state is written as 
\begin{equation}
| \Phi_0 \rangle = \prod^{\rm occ}_{m \sigma} c^+_{m \sigma} | 0 \rangle~~~.
\label{eq:06}
\end{equation}
The product includes all occupied Wannier orbitals.

An alternative choice are {\it local} orbitals consisting of the orthonormal atomic orbitals with creation operators $a^+_{\nu \sigma} (I)$. The relation between the two sets are
\begin{equation}
a^+_{\nu \sigma} (I) = \sum^{\rm occ}_{J m} \gamma^{\rm occ}_{\nu m} (J) c^+_{m \sigma} (J) + \sum^{\rm virt}_{J n} \gamma^{\rm virt}_{\nu n} (J) c^+_{n \sigma} (J)
\label{eq:07}
\end{equation}
indicating that the atomic orbitals have components in the occupied as well as in the virtual SCF orbital space. The atomic orbitals $f_m ({\bf r} - {\bf R}_J)$ are usually better localized than the Wannier orbitals denoted by $w_m ({\bf r} - {\bf R}_j)$ \cite{Kohn59,Mazari97}. This is in particular true for metals which have fractionally filled conduction bands and poorly localized Wannier orbitals \cite{Kohn59}. Therefore one would prefer to use the former, because with their help it is simpler to keep electrons apart. This is indeed the starting point of the Local Ansatz method \cite{Stollhoff80,Fulde92,Fulde95_12}. It was applied in an early  computation of the ground-state wavefunction of, e.g., diamond \cite{Kiel82}, BN \cite{Pirovano91} or polyacetylene \cite{Koenig90}. However, the use of atomic orbitals has a significant draw back. Standard quantum chemistry (QC) computer program packages like MOLPRO \cite{MOLPRO} or MOLCAS \cite{MOLCAS} require that electrons are annihilated in {\it orthogonal} occupied SCF orbitals. Note that this orthogonality requirement does not hold for the creation of electrons in virtual space. Destroying electrons in atomic orbitals implies annihilation in {\it nonorthogonal} states since due to (\ref{eq:07}) only the part of $a^+_{\nu \sigma}$ in the occupied space is annihilated. The development of a QC program package which allows for the annihilation of electrons from nonorthogonal orbitals would be a major advancement in electronic structure calculations. For the above reason $|S)$ has been computed for a large number of compounds by using Wannier orbitals instead of atomic orbitals. For a survey see, e.g., \cite{Paulus06} or \cite{PaulusStoll}.

\section{Impurity embedding methods}
\label{ImpEmbMeth}

Another way of dealing with the EW problem are embedded impurity or cluster theories like the Dynamical Mean Field Theory (DMFT)~\cite{Georges96} or the Density Matrix Embedding Theory (DMET). The DMFT is based on Green's functions. Computations with the latter use connected diagrams only. Similarly, cumulants use connected contractions only, when matrix elements are calculated. In distinction to DMFT the DMET is based on using wavefunctions. Here electrons are treated post-SCF on one site (or cluster), only. The latter is termed fragment. The remaining part, i.e., the embedding bath is treated on a SCF level. Therefore an exponential increase in the number of configurations takes place only for the electrons on the fragment. Embedded impurity theories do not and cannot make any statements about a wavefunction in which all sites are treated on equivalent level. Instead, the embedding procedure is repeated for each site or cluster independently. Double counting of correlation contributions, e.g., to the binding energy is avoided by an assumption about the subdivision of energy contributions  (see below).

The DMET, on which we will concentrate in the following, describes interacting electron systems with orthonormal atomic orbitals. First we discuss the active space within which the correlations are described. For an illustration consider a lattice with $M = 1$ orbitals per site and a half filled SCF band. A selected site $I$ has $2^{2M} = 4$ possible configurations $| \nu (I) \rangle$. They are $| 0 \rangle$, $a^+_\uparrow (I) | 0 \rangle$, $a^+_\downarrow (I) | 0 \rangle$ and $a^+_\uparrow (I) a^+_\downarrow (I) | 0 \rangle$ and define a four dimensional Fock space referring to site $I$. Alternatively, we can characterize the four configurations by four operators $O_\nu (I)$ $(\nu = 1, \dots, 4)$.
\begin{eqnarray}
O_1 (I) & = &\left( 1 - n_\uparrow (I) \right) \left( 1 - n_\downarrow (I) \right) \qquad ; \quad O_3 (I) = n_\downarrow (I) \left( 1 - n_\uparrow (I) \right) \nonumber \\
O_2 (I) & = & n_\uparrow (I) \left( 1 - n_\downarrow (I) \right) \qquad\qquad ~; \quad O_4 (I) = n_\uparrow (I) n_\downarrow (I)
\label{eq:08}
\end{eqnarray}
with the number operator $n_\sigma (I) = a^+_\sigma (I) a_\sigma (I)$. The operators $O_\nu (I)$ act on $| \Phi_0 \rangle$ which serves as reference state for the operators spanning the Liouville space (see (\ref{eq:02})). It is
\begin{equation}
\sum^4_{\nu = 1} O_\nu (I) = 1
\label{eq:09}
\end{equation}
and therefore the following identity holds
\begin{eqnarray}
| \Phi_0 \rangle & = & \sum^n_{\nu = 1} O_\nu (I) | \Phi_0 \rangle \nonumber \\
& = & \sum_\nu | \nu (I) \rangle | B_\nu (I) \rangle~~~.
\label{eq:10}
\end{eqnarray}
The operator $O_\nu (I)$ selects from all the configurations contained in the Slater determinant $| \Phi_0 \rangle$ those, in which site $I$ is in the configuration $| \nu (I) \rangle$. The remaining products of operators formed by the $c^+_{m \sigma} (J)$ define the bath $|B_\nu (I) \rangle$. In $| \Phi_0 \rangle$ the four configurations $| \nu (I) \rangle$ have equal weight. Correlations change these weights. They partially suppress configurations with large electron repulsions and enhance those with low interaction energy. In the special case where the interaction is reduced to an on-site local repulsion U (Gutzwiller-Hubbard Hamiltonian) the weight of double occupancy of site $I$ is reduced. Therefore the correlated ground state is written as
\begin{eqnarray}
| \psi_0 (I) \rangle & = & \sum_\nu \lambda_\nu (I)~ | \nu (I) \rangle | B_\nu (I) \rangle \nonumber\\
& = & \sum_\nu \lambda_\nu (I)~ O_\nu (I) | \Phi_0 \rangle~~~.
\label{eq:11}
\end{eqnarray}
The first line has the form of a Schmidt decomposition used quite frequently in connection with specifying entanglements \cite{Peschel12}. Note that the $|\nu (I) \rangle$ as well as the $| B_\nu (I) \rangle$ do not have a fixed particle number and are therefore vectors in Fock space, while $| \psi_0 (I) \rangle$ is a vector in Hilbert space. The second line expresses $| \psi_0 (I) \rangle$ in terms of operators $O_\nu (I)$. They are acting on $| \Phi_0 \rangle$, and represent a vector in Liouville space. Both representations are equivalent. When we consider $| \Phi_0  \rangle$ as a vacuum state, the operators $( 1- \lambda_\nu (I))~ O_\nu (I)$ describe fluctuations out of the vacuum state. When instead of a single impurity site $I$ we would treat all sites equivalently, then $\sum_\nu ( 1 - \lambda_\nu (I))~ O_\nu (I)$ would be equal to $S_I$ in Eq. (\ref{eq:05}).

The $\lambda_\nu (I)$ can be determined by standard QC methods. This is done for a specific example in the next Section. For a system with equivalent sites it must be ensured that the $\lambda_\nu (I) \neq 1$ do not change the average electron number at site $I$, i.e., $\langle \psi_0 (I) |(n_\uparrow + n_\downarrow)| \psi_0 (I) \rangle = 1$. This subsidiary condition can be easily fulfilled by eliminating single-particle fluctuations. It should be pointed out that the active space of the DMET or DET is identically the same as used in the Local Ansatz method \cite{Stollhoff80, Fulde92,Fulde95_12}. This is obvious for the example considered here where the Local Ansatz uses the same $O_\nu (I)$ \cite{Fulde95_12,Stollhoff80}. Yet it holds also for more general cases. 

The concept of the DMET can, of course, be generalized to $M$ different orbitals per site or fragment $I$. The different configurations do then depend on  the number of electrons $L$ at the fragment, with $0 \leq L \leq 2M$. For each configuration $k$ with fixed number $L$ the representation holds
\begin{equation}
\left| C^L_k (I) \right> = \prod^L_{\nu = 1} a^+_{m_\nu \sigma_\nu} (I) | 0 \rangle~~~.
\label{eq:12}
\end{equation}
Their number depends on the total angular momentum and spin operators, i.e., ${\bf L}^2$, ${\bf S}^2$, $L_z$ and $S_z$. Similarly as in (\ref{eq:08}) we can define operators $O^L_\nu (I)$ which, when applied on $| \Phi_0 \rangle$ filter out all parts of the Slater determinant in which the electrons on fragment $I$ are in the configuration $\left| C^L_k (I) \right>$,
\begin{eqnarray}
| \psi_0 \rangle & = & \sum^{2M}_L \sum_{k(L)} \lambda^L_{k(L)} (I) \left| C^L_{k(L)} (I) \right> \left| B^L_{k(L)} (I) \right>\nonumber \\
& = & \sum^{2M}_L \sum_{k(L)} \lambda^L_{k(L)} (I) O^L_{k(L)} (I) \left| \Phi_0 \right>~~~.
\label{eq:13}
\end{eqnarray}
For an example see Appendix F in \cite{Fulde95_12}.
This form can be again interpreted as a Schmidt decomposition or as a Local Ansatz for operators in Liouville space. We do not want to go into further details, since they are irrelevant for the EW problem discussed here.

\section{From fragment embedding to cumulants and increments}
\label{ImpBeddingCumul}

As pointed out above the embedded fragment wavefunction (\ref{eq:11}) or (\ref{eq:13}) does not face an EW problem, because only a small number of electrons are correlated. Therefore the question arises how to go over from this wavefunction to one in which all sites are equivalent. 

This will tell us how the EW problem is avoided. In order to study this process we consider specifically the $H_{10}$ ring discussed in \cite{Knizia13,Wouters16}. It allows also for a detailed comparism of the DMET with the method of increments \cite{Stoll92,Stoll92a} resulting from Eqs. (\ref{eq:04}) and (\ref{eq:05}).

All following calculations were performed with the MOLPRO ab-initio suite of programs \cite{MOLPRO,Werner12}, using the minimal basis set of Ref. \cite{Knizia13} if not mentioned otherwise. Both DMET and the incremental approach rely on localized orbitals. For a given site ($H$ atom), such orbitals were generated by projecting the corresponding atomic orbital (AO) onto the occupied and virtual SCF space of the $H_{10}$ ring, respectively. This directly corresponds to the DMET active space definition in Ref. \cite{Knizia13}, for a fragment size of just one $H$ atom (which we adopt in the following). Within the DMET method, correlation effects are considered for single fragments individually. All fragments are identical, and the 2-orbital active space of a fragment leads to 3 singlet coupled configuration state functions (CSFs) only. Therefore a single 3-by-3 configuration interaction (CI) calculation is sufficient here. Within the Local Ansatz (LA) \cite{Fulde95_12,Stollhoff80}, on-site excitations are generated,  by applying the operators $O_4$ of Eq. (\ref{eq:08}) to the SCF reference. The latter operators cause double excitations from the occupied part of an AO to the virtual part of the same AO, i.e., they are equivalent to double excitations in the above 2-orbital active spaces. Including such excitations at all atoms
simultaneously, leads to a non-orthogonal CI problem. Within the incremental approach, construction of localized orbitals by projection of AOs as above is also possible (and indeed has been suggested many years ago \cite{Stoll96}), but the incremental expansion is usually based on orthogonal orbitals (at least within the occupied space). This can be achieved by projecting AOs at every second $H$ atom of the ring, with subsequent symmetrical orthogonalization. In the following, this set of orbitals is used if not mentioned otherwise.

\begin{figure}[h]
\includegraphics[width=0.6\textwidth,angle=-90]{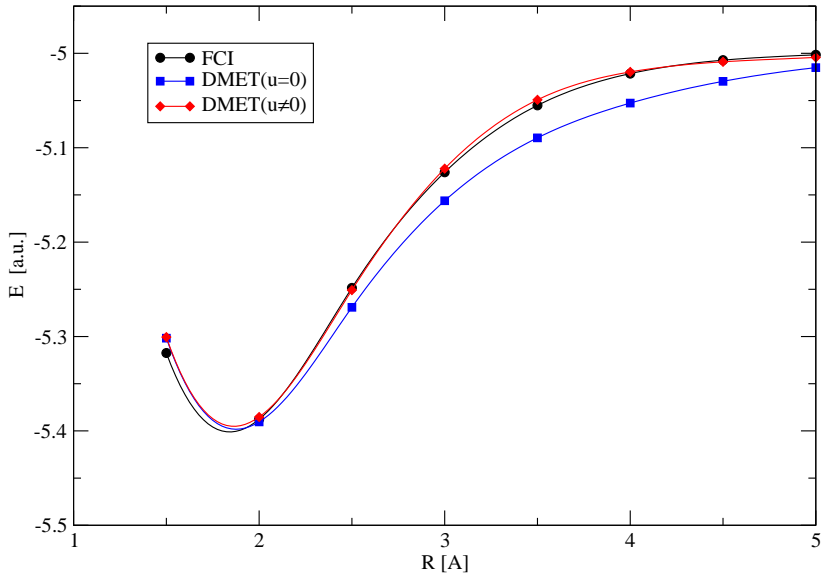}
\caption{Total energy of $H_{10}$ ring (in Hartree units), as a function of the ring radius $R$ (in Angstrom). Full CI (FCI) calculations are compared to calculations using density matrix embedding theory (DMET) with and without additional correlation potential $u$.}
\label{fig:S01}
\end{figure}

Let us begin with a discussion of DMET results for the $H_{10}$ ring, in comparison with corresponding full CI (FCI) results (Fig.\ref{fig:S01}). For the DMET evaluation of the total energy of $H_{10}$, the variational energy of the above-mentioned 3-by-3 CI calculation is not directly used, since this would lead to double counting of energy contributions when summing over all fragments. Instead, one- and two-electron integrals in the basis of symmetrically orthogonalized AOs are multiplied by corresponding one- and two-particle density matrices and by additional weight factors. These weight factors are just the number of fragment AOs within a given integral, divided by the total number of AOs in the integral. Adding up the energies so obtained for all fragments yields the DMET energy for the system as a whole \cite{Wouters16}. This is the important step by which the EW problem is avoided! As shown by the blue curve in Fig.\ref{fig:S01}, it already yields a semi-quantitative approximation to the FCI potential curve of the $H_{10}$ ring. A further improvement can be achieved by adding a correlation potential $u$ at the bath atoms surrounding a given fragment. With this potential, equal charges for all the atoms of the $H_{10}$ ring are maintained. As seen from the red curve in Fig.~\ref{fig:S01}, the resulting potential curve for $H_{10}$ is in excellent quantitative agreement with the reference FCI curve. 

From the very design of the DMET method a correct dissociation of the $H_{10}$ ring into separated single-atom fragments is achieved. However, the excellent performance of DMET for the binding energy at the equilibrium geometry is unexpected. In order to get more insight here, we compared the individual energy contributions mentioned above to corresponding ones from the reference FCI calculation, for which a similar splitting of the energy contributions can be done as in DMET. It turns out that at $R = 2.0 \AA$, the DMET calculation yields a difference of 0.1 a.u. from the reference FCI, for the intra-fragment contributions to the total energy of the $H_{10}$ ring. This deviation is significantly larger than the deviation of the final energies (involving fragment-bath contributions in addition to the intra-fragment ones). Thus, error cancellation between fragment and (less accurate) bath terms seems to be instrumental for DMET here. This means that the high accuracy near the equilibrium bond length may possibly be partially fortuitous. Another hint comes from a CEPA-0 calculation with the Local Ansatz (LA). Since a standard CEPA-0 calculation is in good agreement with the FCI one at $R = 2.0 \AA$, the LA result should reflect merits/shortcomings of the simple on-site excitations. Actually, the LA wavefunction and likewise $\sum\limits^{10}_{I=1} (H|S_I)$ contain on-site excitations for all sites of the $H_{10}$ ring simultaneously and should be superior, therefore, to the DMET ansatz (where the coupling between excitations at different sites is not taken into account). However, it leads to an energy of $-5.32$ a.u. at $R = 2.0 \AA$, which is significantly less accurate than DMET.

\begin{figure}[h]
\includegraphics[width=0.6\textwidth,angle=-90]{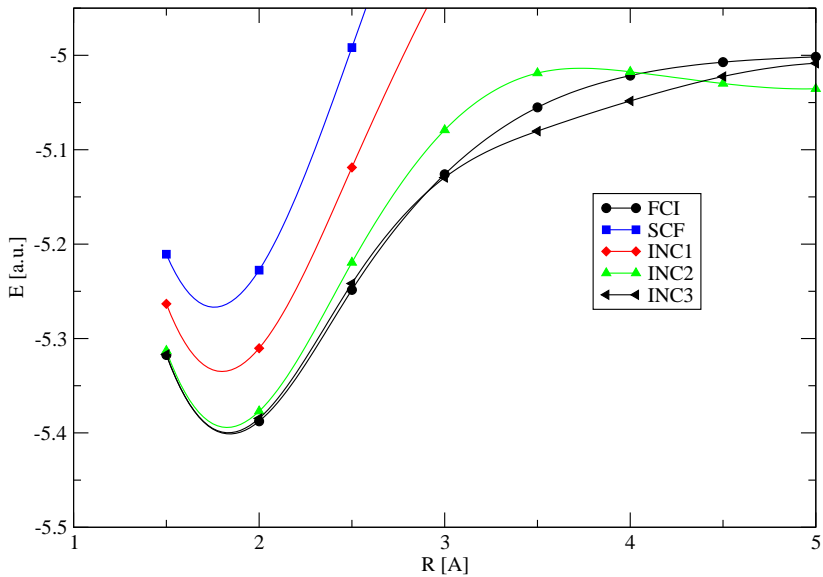}
\caption{Total energy of $H_{10}$ ring (in Hartree units), as a function of the ring radius $R$ (in Angstrom). SCF and full CI (FCI) calculations are compared to calculations using the incremental approach up to $n$-body level (INC$n$). In the latter case, the SCF reference is stepwise improved by subsystem FCI calculations with an active space including up to $n$ localized occupied SCF orbitals (see text) together with the full virtual SCF space.}
\label{fig:S02}
\end{figure}

Let us now discuss incremental approaches. In the standard incremental scheme, the number of correlated localized occupied orbitals $\phi_i$ is systematically enlarged, i.e., at the one-body level the $\phi_i$ are correlated one at a time, at the two-body level pairs $\phi_i$, $\phi_j$ are correlated simultaneously, while excitations into the whole virtual space are possible at all stages. Using the set of localized orbitals described above, we obtain the results shown in Fig.~\ref{fig:S02}. The convergence of the incremental expansion is rather poor at large distances. This is not surprising, since the expansion starts from the restricted HF (RHF) energy, which is increasingly poor for large $R$, so that the incremental corrections become huge. Even at the 3-body level, the accuracy is not really better than with DMET at $u = 0$. The situation is more favourable around the equilibrium $R$. Still, the 1-body level yields only about half of the correlation effect. This was to be expected since we sum up contributions from the correlation of 5 orthogonalized localized occupied orbitals, instead of 10 fragments as in DMET. However, the 2-body and, particularly, the 3-body approximation give a very good account of the potential curve of the $H_{10}$ ring near the equilibrium $R$. Note that the number of 2- and 3-body terms needed in the incremental expansion is quite small, since only nearest-neighbour contributions are non-negligible. Still, the number of correlated orbitals is definitely larger than needed in DMET for the same accuracy.

\begin{figure}[h]
\includegraphics[width=0.6\textwidth,angle=-90]{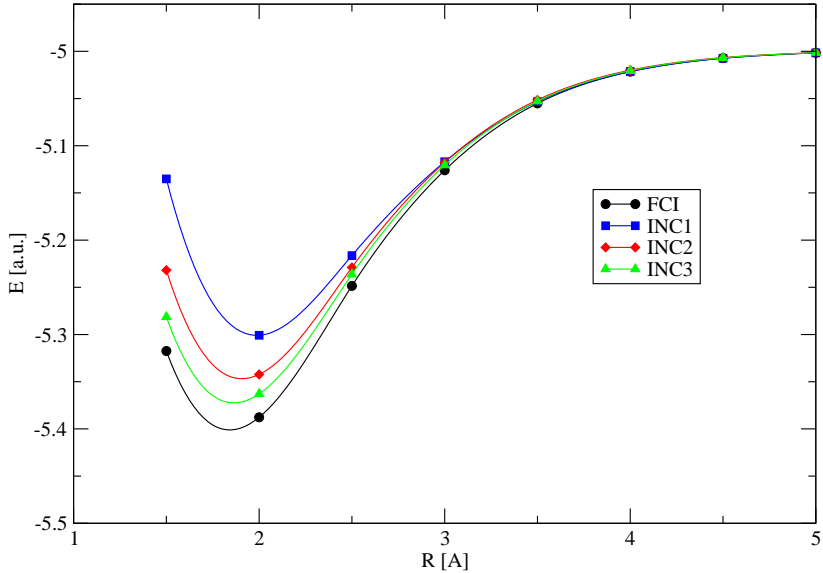}
\caption{Total energy of $H_{10}$ ring (in Hartree units), as a function of the ring radius $R$ (in Angstrom). Full CI (FCI) calculations are compared with calculations using the incremental approach up to $n$-body level (INC$n$). In the latter case, a strongly localized Slater determinant (see text) is stepwise improved by subsystem CASSCF calculations with an active space including up to 2$n$ localized orbitals.}
\label{fig:S03}
\end{figure}

\begin{figure}[h]
\includegraphics[width=0.6\textwidth,angle=-90]{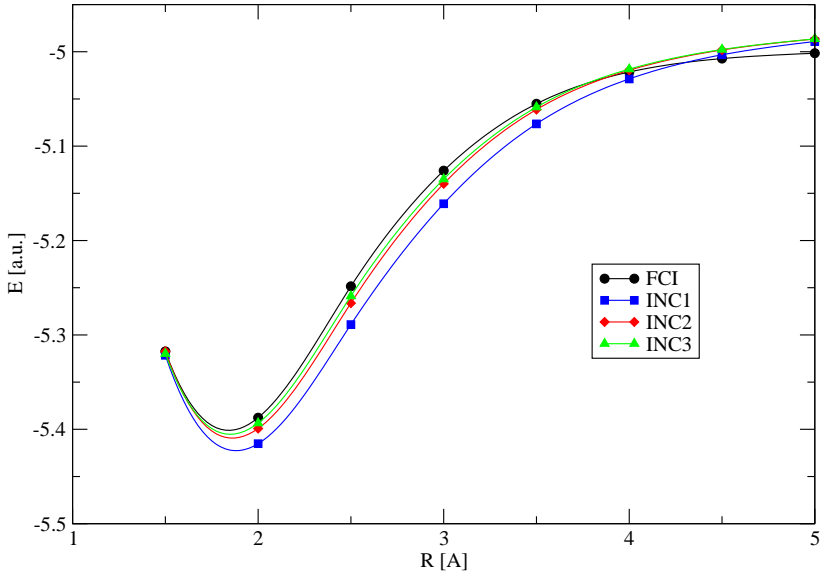}
\caption{Total energy of $H_{10}$ ring (in Hartree units), as a function of the ring radius $R$ (in Angstrom). Full CI (FCI) calculations are compared to calculations using the incremental approach up to $n$-body level (INC$n$). In the latter case, a strongly localized Slater determinant (see text) is stepwise improved by subsystem SCF and CASSCF calculations with an active space including up to $n$ and 2$n$ localized orbitals, respectively. The final incremental expansion is done for the correlation energy only.}
\label{fig:S04}
\end{figure}

The question to be discussed now is: Can one gain insights from DMET for the design of incremental expansions, in order to improve the behaviour for the asymptotic region of the potential curve and/or to reduce the computational effort? The first point concerns entanglement. The RHF wavefunction is strongly entangled for large $R$, and high-level increments
are needed for disentanglement. However, it is not necessary for an incremental expansion to start from RHF. A natural extension would be a start from an unrestricted HF (UHF) wavefunction. It is excluded here for technical reasons related to MOLPRO. But as first shown for metals by Paulus and co-workers \cite{Paulus04,PaulusB04}, a localized model wavefunction can be helpful as a starting point. In the case of the $H_{10}$ ring, a Slater determinant composed of localized 2-atom bond orbitals, i.e., of (normalized) linear combinations $\phi_i + \phi_{i+1}$, $i = 1, 3, 5$ meets the purpose. The latter are not perfect for large $R$, of course, but disentanglement can be easily achieved already at the one-body level, by adding the antibonding linear combinations $\phi_i - \phi_{i+1}$ to the active space. Indeed, as seen from Fig.~\ref{fig:S03}, the asymptotic region is dramatically improved, and good agreement with FCI is already achieved at the one-body level. Note that for the curves of Fig.~\ref{fig:S03}, the active spaces were handled as in DMET, i.e., the virtual space was not included in full as for the incremental expansion above, but restricted to $n$ localized orbitals for $n$-body increments. On the other hand, orbital optimization within the active space turned out to be essential. As a drawback, however, the convergence of the incremental expansion in the bonding region deteriorates. Here, the starting point in terms of localized 2-atom bond orbitals is less appropriate. Still, there is a remedy: one can reduce the magnitude of the energy piece obtained from the incremental expansion, by applying the expansion to the correlation energy only, i.e., not including the improvement of the SCF part. As shown in Fig.~\ref{fig:S04}, this leads to potential-energy curves which are semi-quantitatively correct over the full range of $R$ values, already at the one-body level. Thus, one has (nearly) reached the aim set by DMET, regarding both simplicity and accuracy.

\begin{figure}[h]
\includegraphics[width=0.6\textwidth,angle=-90]{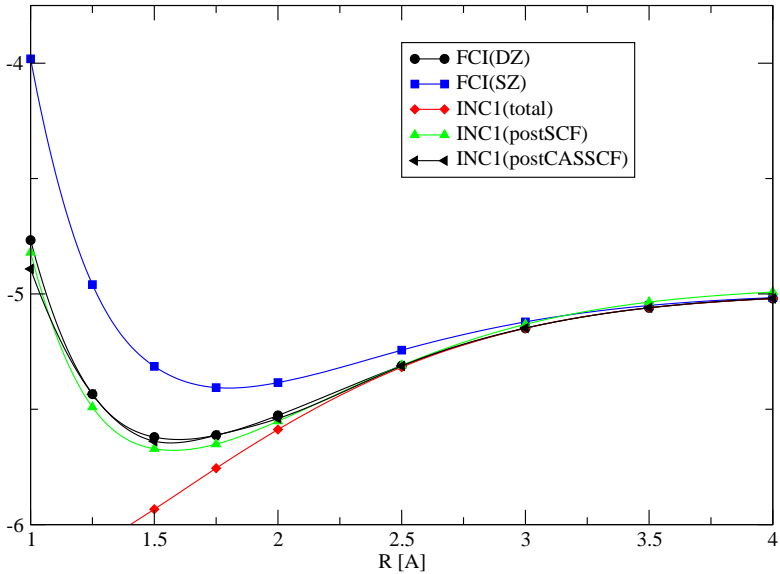}
\caption{Total energy of $H_{10}$ ring (in Hartree units), as a function of the ring radius $R$ (in Angstrom). Full CI (FCI) calculations with two different basis sets (single-zeta (SZ) and double-zeta (DZ)) are compared to calculations using an incremental approach for basis-set enlargement (SZ $\to$ DZ), at the one-body level (INC1). In the latter case, the total, post-SCF, or post-CASSCF pieces of the incremental expansion are used in full, or combined with the DZ SCF and DZ CASSCF energies, respectively.}
\label{fig:S05}
\end{figure}

It is clear that FCI calculations with minimal basis sets are mostly of academic interest only. Therefore, one might ask how to improve on the DMET and incremental results discussed above. The latter approach could have advantages here, since the number of (occupied) orbitals to be correlated does not increase with the size of the basis set (in contrast to the number of AOs which is relevant for DMET). Fig.~\ref{fig:S05} gives results for an incremental scheme of basis-set enlargement, at the one-body level (i.e., enlargement of the $H$ basis set
from minimal to cc-pVDZ \cite{Dunning89} at a single atom only). The basis-set change influences the total energy at all theoretical levels (e.g., SCF, CASSCF, FCI). For the change of the total energy, the incremental expansion up to one-body terms (multiplying the basis-set change at a single atom by the number of atoms in the $H_{10}$ ring) is clearly insufficient, as seen from Fig.~\ref{fig:S05}. However, this approximation works surprisingly well, when applied to basis-set changes at the post-SCF or post-CASSCF levels (in combination with the DZ SCF or DZ CASSCF energy, respectively).

\section{Conclusions and Summary}
\label{Concl_Summary}

Electronic structure calculation for solids have to deal with the exponential wall problem. A natural way is to start from a self-consistent field solution for the ground state and to describe the ground state of the correlated many-electron system by the operators which generate the correlation hole of the electrons. Stated differently, the ground-state wavefunction of the correlated electron system is described by a vector in operator- or Liouville space rather than in Hilbert space. In this space a cumulant metric has to be applied. In other words, whenever matrix elements with correlation generating operators are evaluated only connected contractions of operators have to be taken into account. Disconnected contractions correspond to factorizations of matrix elements and drop out.  This is analogous to Green's functions, where also only connected diagrams have to be taken into account. The use of the cumulant metric eliminates the EW problem.

Another approach to avoid the EW problem is using impurity or fragment embedding. The dynamical mean-field theory (DMFT) \cite{Georges96} is the best known example. As a frequency dependent coherent potential method it is based on Green's function. The EW problem does not arise in this context. But  one would expect that it shows up in the DMET and DET when the wavefunction of embedded {\it impurities} or {\it fragments} is calculated. The basis consists here of orthonormal atomic orbitals and by means of a Schmidt decomposition the system is partitioned into fragment and bath. Only the fragment is treated on a post-SCF level while the bath remains on a SCF level  \cite{Knizia13, Bulik14}.

The active space used hereby is identical with that of the Local Ansatz (LA). In distinction to the DMET which uses Schmidt decompositions, the LA is formulated in terms of projection and partitioning of the Liouville space. It is also based on using connected contractions only and avoids this way the EW problem. In the DMET the EW problem is circumvented by a special extrapolation from the results for largely independent fractions, coupled via correlation potentials, to those of many fragments. This extrapolation is scrutinized here. By studying a $H_{10}$ ring for different radii we have related the DMET results with those based on the method of increments which is related to the cumulant approach. What remains to be done is to relate the cumulant or Liouville approach to tensor networks \cite{Oros14} and to matrix product states \cite{Schollwoeck05,Verstraete08}.

\newpage



\begin{center}  
\Large{References} 
\end{center}

\begin{thebibliography}{50}
\bibitem{Kohn99} W. Kohn, Rev. Mod. Phys. {\bf 71}, 1253 (1999)
\bibitem{Hohenberg64} P. Hohenberg, and W. Kohn, Phys. Rev. B {\bf 136}, 864 (1964)
\bibitem{Kohn65} W. Kohn, and L. Sham, Phys. Rev. {\bf 140} (4A), 1133 (1965)
\bibitem{Cizek69} J. Cizek, Adv. Chem. Phys. {\bf 14}, 35 (1969)
\bibitem{Kuemmel78} H. K\"ummel, K. H. L\"uhrmann, and J. G. Zabolitzky, Phys. Reports {\bf 36}, 1 (1978)
\bibitem{Bishop91} R. F. Bishop, Theor. Chim. Acta {\bf 80}, 95 (1991)
\bibitem{Sun99} J.-Q. Sun, and R. Bartlett, {\it Correlation and Localization}, Topics in Current Chemistry, Vol. {\bf 203}, ed. by P. Surjan et al. (Springer Berlin)  (1999)
\bibitem{Manby11} {\it Accurate Condensed-Phase Quantum Chemistry}, ed. by F. R. Manby (CRC Press, Boca Raton) (2011) and references therein
\bibitem{Booth13} G.H. Booth, A. Gr\"uneis, G. Kresse, and A. Alavi
Towards, Nature {\bf 493}, 365 (2013)
\bibitem{Sharma15} S. Sharma, and A.Alavi, J. Chem. Phys. {\bf 143}, 102815 (2015)
\bibitem{Kersten16} J. A. F. Kersten, G. H. Booth, and A. Alavi, J. Chem. Phys. {\bf 145}, 054117 (2016)
\bibitem{Georges96} A. Georges, G. Kotliar, W. Krauth, and W. Rosenberg, Rev. Mod. Phys. {\bf 68}, 13 (1996)
\bibitem{Knizia13} G. Knizia, and K.-L. Chan, J. Chem. Theory Comput. {\bf 9}, 1428 (2013)
\bibitem{Bulik14} I. W. Bulik, W. Chen, and G. E. Scuseria, J. Chem. Phys. {\bf 141}, 054113 (2014)
\bibitem{Fulde95_12}
Fulde, P., {\it Correlated Electrons in Quantum Matter} (World Scientific Publ., Singapore) (2012)
\bibitem{Kladko98} K. Kladko, and P. Fulde, Int. J. Quantum Chem. {\bf 66}, 377 (1998)
\bibitem{Fulde16} P. Fulde, Nature Phys. {\bf 12}, 106 (2016)
\bibitem{Schork92} T. Schork, and P. Fulde, J. Chem. Phys. {\bf 97}, 9195 (1992)
\bibitem{Loewdin63} P. O. L\"owdin, J. Mol. Spectrosc. {\bf 10}, 12 (1963); ibid {\bf 13}, 326 (1964); ibid {\bf 14}, 112 (1964)
\bibitem{Loewdin86} P. O. L\"owdin, Int. J. Quantum Chem. {\bf 21}, 29 (1982); ibid {\bf 29}, 1651 (1986)
\bibitem{Stollhoff80} G. Stollhoff, and P. Fulde, J. Chem. Phys. {\bf 73}, 4548 (1980)
\bibitem{Fulde92} P. Fulde, and G. Stollhoff, Int. J. Quantum Chem. {\bf 42}, 103 (1992)
\bibitem{Mayer40} E. Mayer, and M. G. Mayer, {\it Statistical Mechanics} (Wiley, New York) (1940)
\bibitem{Kubo66} R. Kubo, Rep. Progr. Phys. {\bf 29}, 255 (1966)
\bibitem{Stoll92} H. Stoll, Phys. Rev. B {\bf 46}, 6700 (1992)
\bibitem{Stoll92a} H. Stoll, J. Chem. Phys. {\bf 97}, 84 (1992)
\bibitem{Paulus06} B. Paulus, Phys. Rev. {\bf 428}, 1 (2006)
\bibitem{Kohn59} W. Kohn, Phys. Rev. {\bf 115}, 809 (1959)
\bibitem{Mazari97} N. Mazari, and D. Vanderbilt, Phys. Rev. B {\bf 56}, 12847 (1997)
\bibitem{Kiel82} B. Kiel, G. Stollhoff, C. Weigel, P. Fulde, and H. Stoll, Z. Phys. B {\bf 46}, 1 (1982)
\bibitem{Pirovano91} M. V. Ganduglia-Pirovano, and G. Stollhoff, Phys. Rev. B {\bf 44}, 3526 (1991) 
\bibitem{Koenig90} G. K\"onig, and G. Stollhoff, Phys. Rev. Lett. {\bf 65}, 1239 (1990)
\bibitem{MOLPRO} MOLPRO, version 2015.1, is a package of ab initio programs written by H.-J. Werner and P. J. Knowles, G. Knizia, F. R. Manby, M. Sch\"utz and others, see http://www.molpro.net
\bibitem{MOLCAS} MOLCAS6, Univ. Lund, Sweden, Univ. of Lund, Dept. of Theoret. Chemistry (2004)
\bibitem{PaulusStoll} B. Paulus, and H. Stoll in: {\it Accurate Condensed-Phase Quantum Chemistry} ed. by F. R. Manby, p. 57, (CRC Press, Boca Raton) (2011)
\bibitem{Peschel12} I. Peschel, Braz. J. Phys. {\bf 42}, 267 (2012)
\bibitem{Wouters16} S. Wouters, C.A. Jim\'enez-Hoyos, Q. Sun, and G. K.-L. Chan, J. Chem. Theory Comput. {\bf 12}, 2706 (2016)
\bibitem{Werner12} H.-J. Werner, P. J. Knowles, G. Knizia, F. R. Manby, and M. Sch\"utz, WIREs Comput. Mol. Sci. {\bf 2}, 242 (2012)
\bibitem{Stoll96} H. Stoll, Ann. Physik {\bf 508}, 355 (1996)
\bibitem{Paulus04} B. Paulus, and K. Ro\'sciszewski, Chem. Phys. Lett. {\bf 394}, 96 (2004)
\bibitem{PaulusB04} B. Paulus, K. Ro\'sciszewski, N. Gaston, P. Schwerdtfeger, and H. Stoll, Phys. Rev. B {\bf 70}, 165106 (2004)
\bibitem{Dunning89} T. H. Dunning, Jr., J. Chem. Phys. {\bf 90}, 1007 (1989)
\bibitem{Oros14} R. Or\'os, Annals of Physics {\bf 349}, 117 (2014)
\bibitem{Schollwoeck05} U. Schollw\"ock, Rev. Mod. Phys. {\bf 77}, 259 (2005)
\bibitem{Verstraete08} F. Verstraete, V. Murg, and J. I. Cirac, Adv. in Phys. {\bf 57}, 143 (2008)

\end{thebibliography}
\end{document}